\pdfoutput=1

\documentclass[sigconf,screen]{acmart}

\AtBeginDocument{%
  }

\setcopyright{acmcopyright}
\copyrightyear{2023}
\acmYear{2023}
\acmDOI{XXXXXXX.XXXXXXX}

\acmConference[]{}{}{}
\acmPrice{15.00}
\acmISBN{978-1-4503-XXXX-X/18/06}

\usepackage{amsmath,amssymb}  %
\usepackage{multirow}  %
\usepackage{listings}
\usepackage{enumitem}
\usepackage{xcolor}  %

\colorlet{punct}{red!60!black}
\definecolor{background}{HTML}{F7F7F7}
\definecolor{delim}{RGB}{20,105,176}
\colorlet{cmnt}{magenta!60!black}

\lstdefinelanguage{json}{
    basicstyle=\normalfont\ttfamily,
    stepnumber=1,
    numbersep=8pt,
    showstringspaces=false,
    breaklines=true,
    frame=lines,
    backgroundcolor=\color{background},
    morecomment=[s]{/*}{*/},
    commentstyle=\color{cmnt}\ttfamily,
    literate=
     *{:}{{{\color{punct}{:}}}}{1}
      {,}{{{\color{punct}{,}}}}{1}
      {\{}{{{\color{delim}{\{}}}}{1}
      {\}}{{{\color{delim}{\}}}}}{1}
      {[}{{{\color{delim}{[}}}}{1}
      {]}{{{\color{delim}{]}}}}{1},
}

\begin{document}

\title[unarXive 2022: All arXiv Publications Pre-Processed for NLP, Including Structured Full-Text and Citation Network]{unarXive 2022: All arXiv Publications Pre-Processed for NLP, Including Structured Full-Text and Citation Network}

\author{Tarek Saier}
\orcid{0000-0001-5028-0109}
\email{tarek.saier@kit.edu}
\affiliation{%
  \institution{Karlsruhe Institute of Technology} %
  \streetaddress{Kaiserstr. 89}
  \city{Karlsruhe}
  \country{Germany}
  \postcode{76133}
}

\author{Johan Krause}
\orcid{0000-0002-5080-3587}
\email{johan.krause@student.kit.edu}
\affiliation{%
  \institution{Karlsruhe Institute of Technology} %
  \streetaddress{Kaiserstr. 89}
  \city{Karlsruhe}
  \country{Germany}
  \postcode{76133}
}

\author{Michael F{\"a}rber}
\orcid{0000-0001-5458-8645}
\email{michael.faerber@kit.edu}
\affiliation{%
  \institution{Karlsruhe Institute of Technology} %
  \streetaddress{Kaiserstr. 89}
  \city{Karlsruhe}
  \country{Germany}
  \postcode{76133}
}

\renewcommand{\shortauthors}{Saier et al.}

\begin{abstract}
Large-scale data sets on scholarly publications are the basis for a variety of bibliometric analyses and natural language processing (NLP) applications. Especially data sets derived from publication's \emph{full-text} have recently gained attention.
While several such data sets already exist, we see key shortcomings in terms of their domain and time coverage, citation network completeness, and representation of full-text content.
To address these points, we propose a new version of the data set unarXive.
We base our data processing pipeline and output format on two existing data sets, and improve on each of them. Our resulting data set comprises 1.9 M publications spanning multiple disciplines and 32 years. It furthermore has a more complete citation network than its predecessors and retains a richer representation of document structure as well as non-textual publication content such as mathematical notation. In addition to the data set, we provide ready-to-use training/test data for citation recommendation and IMRaD classification. All data and source code is publicly available at \url{https://github.com/IllDepence/unarXive}.
\end{abstract}



\maketitle

\section{Introduction}

Large data sets derived from the full-texts of academic publications are of ever-increasing importance. Beyond large-scale metadata, which is the basis for bibliometric analyses, research output quantification~\cite{Hirsch2005}, and various applications such as trend detection~\cite{Chen2006}, data sets reflecting the \emph{full-text} content of papers have recently enabled more sophisticated analyses and applications, such as scientific document summarization~\cite{citesum}, claim verification~\cite{wadden2020}, and knowledge graph generation~\cite{luan2018scierc}.

\begin{figure}
  \centering
  \includegraphics[width=\linewidth]{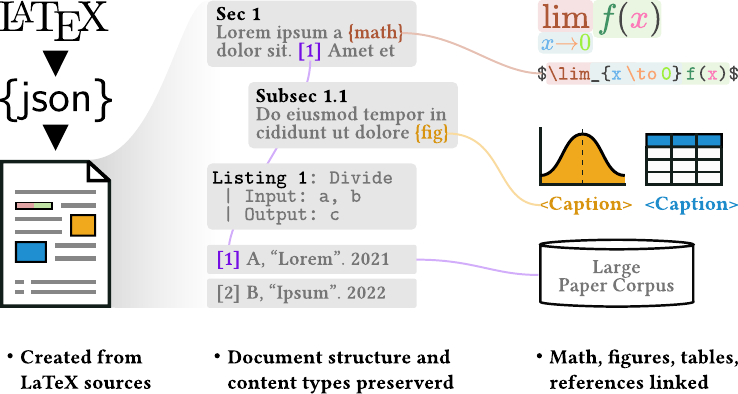}
  \caption{Schematic of our data set.\\
    \footnotesize\normalfont Created from arXiv.org \LaTeX\ sources, our data set preserves document \emph{sctructure} (sections, subsections, ...) and \emph{content types} (paragraphs, listings, ...). In-text positions of mathematical notation, figures, tables and citation markers are linked to \LaTeX\ math content, figure/table captions, and bibliographic references respectively. Bibliographical references are linked to the large paper corpus OpenAlex.
  }
  \label{fig:schema}
\end{figure}

Key aspects of such data sets are (1)~basic measures such as quality, size, and temporal as well as disciplinary coverage, (2)~their citation network, and (3)~handling of non-textual content. (1)~Quality is affected by the source material (e.g. PDF or \LaTeX) and parsing method. (2)~The citation network is important to allow for bibliometric analyses. (3)~Non-textual content such as tables, figures, and mathematical notation often contain important information.

Across these key aspects, we see significant shortcomings in currently available data sets, as shown in Table~\ref{tab:comparison}. For example, (1)~limited size (SciXGen), (2)~omission of a citation network (arXMLiv), and (3)~no or limited handling of mathematical notation (S2ORC, unarXive 2020).

\begin{table*}
  \caption{
    Comparison of large data sets derived from paper full-texts\\
    {\footnotesize\normalfont
     $^\dagger$Cit. network completeness is reported is two ways. ``general'': the whole data set; not directly comparable. ``compare'': for arXiv.org data from 1991--2020; directly comparable.\\
     $^\ddagger$References in the PMC-OAS are partially linked to a mixed set of IDs (PubMed, MEDLINE, DOI)~\cite{Gipp2015}. Therefore there is no single, comprehensive number for its completeness.
    }
   }
  \label{tab:comparison}
  \begin{tabular}{lccccccccc}
    \toprule
    \ & \multicolumn{2}{c}{Source} & \multicolumn{2}{c}{\hphantom{wi}Citation Network$^\dagger$} & \multicolumn{2}{c}{Structured} & \ & \ & \ \\
    Data Set & Data & Format & general & compare & Doc. & Math. & \# Docs & Disciplines & Purpose \\
    \midrule
    CORE~\cite{core} & multiple & PDF & 0\% & - & $\times$ & $\times$ & $>$100 M & various & general NLP \\
    S2ORC (PDF)~\cite{Lo2020} & multiple & PDF & 69.4\% & - & \checkmark & $\times$ & 12 M & various & general NLP \\
    unarXive 2020~\cite{Saier2020} & arXiv.org & \LaTeX & 42.6\% & 42.6\% & $\times$ & $\times$ & 1.2 M & phys., maths, CS & general NLP \\
    \midrule
    S2ORC (\LaTeX)~\cite{Lo2020} & arXiv.org & \LaTeX & 31.1\% & 31.1\% & \checkmark & \checkmark & 1.5 M & phys., maths, CS & general NLP \\
    arXMLiv~\cite{arXMLiv} & arXiv.org & \LaTeX & 0\% & 0\% & \checkmark & \checkmark & 1.6 M & phys., maths, CS & maths linguistics \\
    SciXGen~\cite{chen2021-scixgen} & arXiv.org & \LaTeX & 41.6\% & - & \checkmark & \checkmark & 205 k & CS & text generation \\
    PMC-OAS~\cite{pmc_oas} & PubMed & XML & mixed$^\ddagger$ & - & \checkmark & \checkmark & \textbf{3.3 M} & \textbf{biomedical} & not NLP specific \\
    \textbf{unarXive 2022} (ours) & arXiv.org & \LaTeX & 44.4\% & \textbf{44.4\%} & \textbf{\checkmark} & \textbf{\checkmark} & \textbf{1.9 M} & \textbf{phys., maths, CS} & general NLP \\
    \bottomrule
  \end{tabular}
\end{table*}

To address these issues, we propose a new version of the data set unarXive, %
which comprises 1.9 M publication across several disciplines, includes a more complete citation network than its predecessors, and retains structured mathematical notation as well as table and figure captions (see Figure~\ref{fig:schema}).  %
Apart from the data set itself, we furthermore provide ready-to-use training and test data for two NLP tasks. Overall, we make the following contributions.

\begin{itemize}
    \item We provide a 1.9 M document scholarly data set, containing structured full-text, annotated in-text citations, linked table and figure captions, structured mathematical notation, and a hight quality citation network.
    \item We provide ready-to-use training/test data for the development and evaluation of approaches to two NLP tasks, namely citation recommendation and IMRaD classification.
    \item We distribute our data in accordance to the FAIR principles~\cite{Wilkinson2016} and share our source code freely available under a permissive license.
\end{itemize}

\section{Related Work}

In Table~\ref{tab:comparison} we give an overview of related work.
Excluded are data sets that are either just sets of PDFs, or only contain metadata.

CORE~\cite{core}, while being very large, does not contain a citation network, nor is document structure preserved.
S2ORC (PDF)~\cite{Lo2020} is second in size and, while not directly comparable due to different publications covered, has the most complete citation network. However, mathematical notation is only partially preserved as plaintext.
unarXive 2020~\cite{Saier2020} has the second highest citation network completeness in direct comparison, but lacks structured content.

The bottom part of the table are data sets with both document structure preserved and structured mathematical notation.
S2ORC (\LaTeX)~\cite{Lo2020} is a discontinued\footnote{Last release including the \LaTeX\ subset is 2019-09-28, see \url{https://github.com/allenai/s2orc} (accessed 2023/02/12).} subset of S2ORC and has a limited citation network, 
arXMLiv~\cite{arXMLiv} offers the highest level of structure but no citation network, and 
SciXGen~\cite{chen2021-scixgen} is limited in size.
The PMC-OAS~\cite{pmc_oas} is comparable to unarXive 2022 in size and structure, but has a partial and mixed citation network.

Overall, unarXive 2022 has the most complete citation network as far as direct comparison is possible, preserves document structure as well as structured mathematical notation, and is the largest data set covering physics, mathematics and computer science.

\section{Approach}

We base our data set creation approach in part on S2ORC (\LaTeX) and in part on unarXive 2020. This is motivated as follows.

As shown in Table~\ref{tab:comparison}, the majority of related data sets is based on paper's \LaTeX\ sources---which is less noise-prone than parsing PDFs~\cite{Bast2017}. Among these, S2ORC (\LaTeX) provides well structured full-text content usable for a wide variety of applications (see Section~\ref{sec:applications}), while arXMLiv and SciXGen are optimized for special purposes. We therefore base our structured document representation on S2ORC (\LaTeX). Regarding the citation network, however, unarXive 2020 achieves the most high quality results in direct comparison among existing data sets. We therefore base our citation network creation on unarXive 2020.

Regarding both S2ORC (\LaTeX) and unarXive 2020, we don't just copy, but also improve upon the existing work. To furthermore provide an up-to-date data set, we use as source data all papers on arXiv.org up until the end of 2022.

Conceptually, our overall data set creation process can be broken down into two major steps, namely document parsing and reference linking, In the following these are described in more detail.

\subsection{Document Parsing}
To convert the \LaTeX\ source of a paper into a format that is well suited for NLP applications and analyses, we follow S2ORC (\LaTeX) and unarXive 2020 and perform the following three steps. First, we flatten the paper's \LaTeX\ source into a single \texttt{.tex} document using latexpand.\footnote{See \url{https://ctan.org/pkg/latexpand}.} Next, we use the tool Tralics\footnote{See \url{https://www-sop.inria.fr/marelle/tralics/}.} to convert the \LaTeX\ source into XML. In the last step, we create an easy to handle JSON structure from the XML.

We adapt and extend the JSON structure of S2ORC as shown in Table~\ref{tab:formext}. Adding paper metadata facilitates easier analyses (e.g. for specific or across disciplines). Including information on section numbers and types reflects the document structure more closely (e.g. the nesting structure is not lost). Retaining URLs from embedded links helps with reference linking (see Section~\ref{sec:reflink}).

We mark the position of citation markers, tables, figures, and mathematical notation within the running text, and link citations markers to their references, tables and figures to their captions (i.e., textual surrogates of their content), and mathematical notation to its original \LaTeX\ content.

\begin{table}
  \caption{Extension of S2ORC format}
  \label{tab:formext}
  \begin{tabular}{p{1.5cm}p{2.1cm}p{3.7cm}}
    \toprule
    Entity & S2ORC data & Added data \\
    \midrule
    \textbf{Paper} &
        \begin{minipage}[t]{\linewidth}
            \begin{itemize}[nosep,after=\strut,leftmargin=1mm]
                \item ID
                \item abstract
                \item full-text (list of paragraphs)
                \item bibliographic references
            \end{itemize}
        \end{minipage} &
        \begin{minipage}[t]{\linewidth}
            \begin{itemize}[leftmargin=1mm]
                \item Metadata (title, list of authors, discipline, license, version history)
            \end{itemize}
        \end{minipage}\\
    \textbf{Paragraph} &
        \begin{minipage}[t]{\linewidth}
            \begin{itemize}[leftmargin=1mm]
                \item Section title 
                \item text
            \end{itemize}
        \end{minipage} &
        \begin{minipage}[t]{\linewidth}
            \begin{itemize}[leftmargin=1mm]
                \item Section number 
                \item Section type (e.g. \textit{section}, \textit{subsection}) 
                \item Content type (e.g. \textit{paragraph}, \textit{listing}, \textit{proof})
            \end{itemize}
        \end{minipage}\\
    \textbf{Bib\-li\-o\-gra\-phic reference} &
        \begin{minipage}[t]{\linewidth}
            \begin{itemize}[leftmargin=1mm]
                \item Parsed reference
                \item ID of cited document
            \end{itemize}
        \end{minipage} &
        \begin{minipage}[t]{\linewidth}
            \begin{itemize}[leftmargin=1mm]
                \item Raw reference string
                \item List of contained arXiv IDs
                \item List of embedded links (i.e. URLs of clickable links not rendered as text when viewing the document)
            \end{itemize}
        \end{minipage}\\
  \bottomrule
\end{tabular}
\end{table}

\subsection{Reference Linking}\label{sec:reflink}

To add a citation network to the data set, bibliographical references---which at this point are just raw strings of text---need to be associated with the cited documents they're referencing. We follow the methodology of unarXive 2020 and link references to a large corpus of publication metadata. To do this, references are first parsed to determine the contained information (title, authors, year, venue, etc.), which is then matched against the paper records in the large metadata corpus. For these two steps, we make the following changes and improvements of the unarXive 2020 approach.

\paragraph{Parsing} unarXive 2020 utilizes the tool Neural Parscit~\cite{neuralparscit} for reference parsing and furthermore uses a heuristic procedure to determine identifiers such as DOIs or arXiv IDs found within reference string. We use GROBID~\cite{Lopez2009}, a more commonly used and actively developed tool. Additionally, we extend the identifier determination heuristics to be more robust and versatile by refining matching patterns and extending them to more citation styles.

\paragraph{Matching} unarXive 2020 matches references to paper records in the Microsoft Academic Graph (MAG)~\cite{Sinha2015MAG}, which is no longer publicly available. Instead of the MAG, we use OpenAlex~\cite{openalex}, the MAG's open successor provided by the nonprofit organization OurResearch.\footnote{See \url{https://ourresearch.org/}.} Chosing OpenAlex allows us to also match references to recent papers, which would not be contained in legacy versions of the MAG. Additionally, the fact that OpenAlex paper records contain a variety of identifiers (e.g. DOI and PubMed ID) facilitates combined and comparative analyses of our data with others. Furthermore, OpenAlex has been deemed better suited for bibliographic analyses than the MAG~\cite{openalex-vs-mag}.

\section{Results}

In the following, we first present key statistics of our proposed data set. Following that, we explain how the data set can be used for analyses as well as the development of NLP applications, and introduce training/test data for two NLP tasks. Lastly, we describe how the data set is distributed to facilitate easy adoption by the community of researchers and practitioners.

\subsection{Data Set}

Our data set comprises \emph{1,881,346 papers}, which contain a combined \emph{182,586,547} paragraphs, \emph{63,367,836 references} and \emph{133,744,613 in-text citation markers}. The distribution across disciplines is 57\% physics, 20\% mathematics, 17\% computer science, and a combined 5\% for others. %
We are able to link 28,135,565 references (44.4\%) and 64,547,944 (48.3\%) in-text citation markers to OpenAlex. As shown in Table~\ref{tab:comparison}, this makes our citation network more complete than that of existing data sets.

In Listing~\ref{lst:datasample} we show an excerpt of our document representation for one paper, showcasing the extracted plain text and structured content.

\begin{lstlisting}[language=json,caption=Data example.,label=lst:datasample,breaklines=true,captionpos=b,frame=single,showlines=true,basicstyle=\tiny]
/* - - - - - - - example paper (arXiv:2105.05862) - - - - - - - */
{ "paper_id": "2105.05862",
  "metadata": {...},
  "abstract": {...},
  "body_text": [...],
  "ref_entries": {...},
  "bib_entries": {...} }
/* - - - - - - - one of the sections in body_text - - - - - - - */
{ "section": "Memory wave form",
  "sec_number": "2.1",
  "sec_type": "subsection",
  "content_type": "paragraph",
  "text": "The gauge choice leading us to this solution does not fix
           completely all the gauge freedom and an additional constraint
           should be imposed to leave only the physical degrees of freedom.
           This is done by projecting the source tensor {{formula:7fd88bcd-
           9013-433d-9756-b874472530d9}} into its transverse-traceless (TT)
           components (see for example {{cite:80dbb6c8b9c12f561a8e585faceac5f
           4e104d60d}}). Doing this and without loss of generality, we will
           use the following very well known ansatz for the source term
           proposed in {{cite:bc9a8ca19785627a087ae0c01abe155c22388e16}}\n" }
/* - - - - - - - ref_entries entry for {{formula:7fd88...}} - - - - - - - */
{ "latex": "S_{\\mu \\nu }",
  "type": "formula" }
/* - - - - - - - bib_entries entry for {{cite:80dbb...}} - - - - - - - */
{ "bib_entry_raw": "R. Epstein, The Generation of Gravitational Radiation by Esc
                    aping Supernova Neutrinos, Astrophys. J. 223 (1978) 1037.",
  "contained_links": [
    { "url": "https://doi.org/10.1086/156337",
      "text": "Astrophys. J. 223 (1978) 1037.",
      "start": 87,
      "end": 117 }
  ],
  "ids" {...} }
\end{lstlisting}

\begin{figure}
  \centering
  \includegraphics[width=\linewidth]{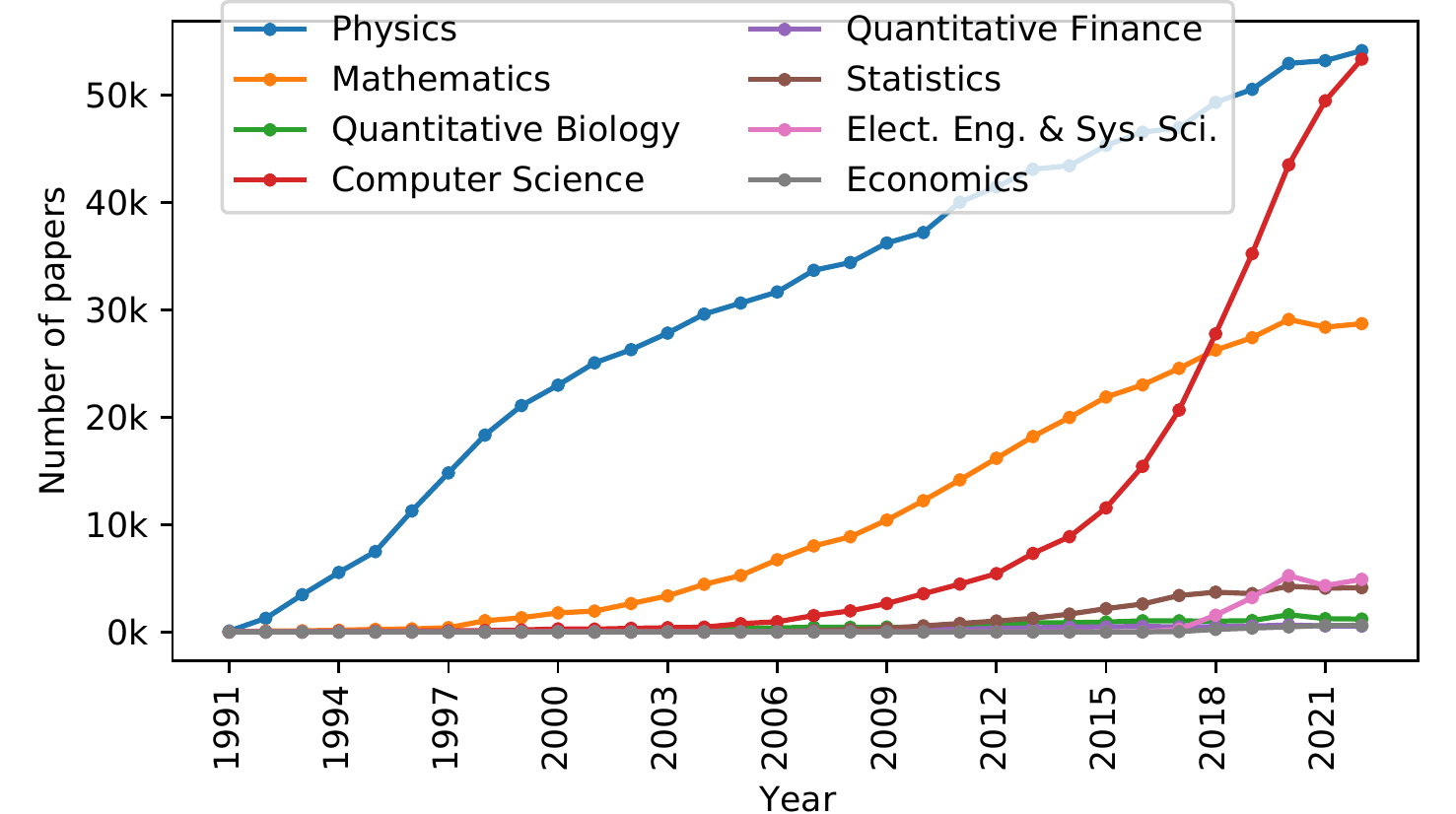}
  \caption{Number of papers per year.}
  \label{fig:numpprs}

  \includegraphics[width=\linewidth]{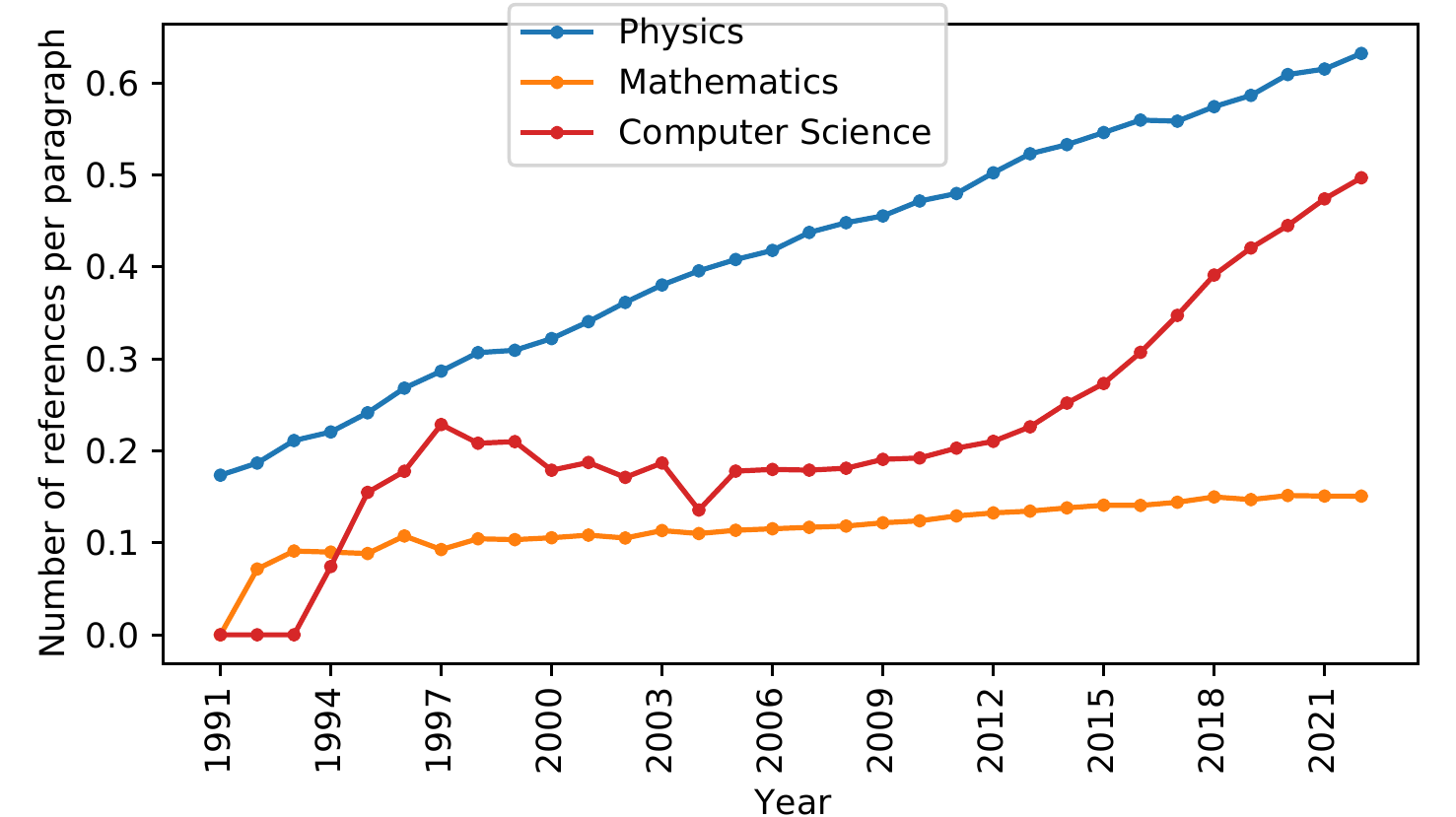}
  \caption{Reference density per year.}
  \label{fig:refdensity}
\end{figure}

In Figure~\ref{fig:numpprs} we show the number of papers across all disciplines over all years covered. We can see that yearly arXiv.org submissions in computer science are likely to surpass those in physics in 2023. As a simple showcase of the use of structured full-text content, we show in Figure~\ref{fig:refdensity} how the average number of bibliographic references per paragraph developed over time for the three major disciplines represented in the data set. Dividing by paragraphs is done to account for variation in paper length. We can see that the density of references is increasing more rapidly in physics and computer science, than it is in mathematics.

\subsection{Applications}\label{sec:applications}

As is evident by the past use of our data set's predecessors unarXive 2020 and S2ORC, large-scale scholarly data sets created with NLP research in mind have broad applicability. Example uses are analyses of citation behavior across languages~\cite{Saier2021} or disciplines~\cite{astrocit} and the development of models for claim verification~\cite{wadden2020}, document retrieval~\cite{multi_objective}, summarization~\cite{citesum}, or information extraction~\cite{viswanathan2021}.

Due to its similarities in structure and contained information, unarXive 2022 is equally suitable for the applications named above. Beyond these, we provide data for two NLP tasks on unarXive 2022, namely content based citation recommendation and IMRaD classification, which are described in the following.

\paragraph{Content Based Citation Recommendation}
Given a piece of text and a citation-marker position, the task of content based citation recommendation entails identifying publications which are suitable to cite in the given text at the given position~\cite{Bhagavatula2018,Faerber2020a}. Large full-text corpora of publications with a citation network provide a rich source for supervision of machine learning (ML) models for this task. That is, human made citations are used as training examples, or for evaluating models in a citation re-prediction setting.
From the premissively licensed papers in our data set we use all in-text citation markers with a linked reference cited at least three times, to allow splitting into train, dev, and test data.
The result is 2.5 M items consisting of (1) a paragraph and citation marker position (model input), and (2) the ID of the cited document (desired model output). %

\paragraph{IMRaD Classification}
Scientific publications are usually structured into sections commonly summarized as ``Introduction, Methods, Results, and Discussion'' (IMRaD). Classifying sections of scientific text into these four classes is done, for example, in fine-grained citation classification. %
Because conventions differ between disciplines, we prepare data for this task for computer science papers only. To aforementioned four classes we add the common ``Related Work'' section as a fifth class. From the premissively licensed computer science papers in our data, we use those that are unambiguously assignable to one of the five classes. The result is 530~k items consisting of (1) the paragraph text (model input), and (2) the class (desired model output). An exemplary application scenario for a model trained on this data is a paper writing assistant that can detect parts in a manuscript, which might be better placed in a different section (e.g. discussion rather than results).

\subsection{Distribution}

Under consideration of the FAIR principles, we chose the following well established distribution channels and licenses for our data set, aforementioned NLP task data, as well as our source code.

\begin{itemize}
    \item The \textbf{data set} is distributed on \texttt{Zenodo}.\\
        $\rightarrow$\url{https://doi.org/10.5281/zenodo.7752615} (open subset)\\
        $\rightarrow$\url{https://doi.org/10.5281/zenodo.7752754} (full)\\
        In accordance with the licensing terms of our source data, we share our data set in two versions.\\
        (1) The subset generated from permissively licensed source data (165 k publications, 9\%) is openly accessible.\\
        (2) The full data set, generated partially from source data under arXiv.org's ``non-exclusive license to distribute,''\footnote{See \url{http://arxiv.org/licenses/nonexclusive-distrib/1.0/}.} is accessible through Zenodo's ``restricted access'' policy, %
making it possible to grant access to the data on request given the intended use is in accordance with the license terms.
    \item The \textbf{NLP task data} is provided on the \texttt{Hugging Face Hub}.\\
        $\rightarrow$\url{https://huggingface.co/datasets/saier/unarXive_citrec}\\
        $\rightarrow$\url{https://huggingface.co/datasets/saier/unarXive_imrad_clf}\\
        This facilitates easy access and use by the NLP community.
    \item The \textbf{source code} for creating the data set is shared on \texttt{GitHub} under the MIT License.\\
        $\rightarrow$\url{https://github.com/IllDepence/unarXive}\\
        Sharing the code openly and permissively licensed allows anyone to freely modify and extend the code to their needs. This makes, for example, integration into other NLP projects such as benchmarks and frameworks possible.
\end{itemize}

\section{Conclusion}
We propose unarXive 2022, a data set generated from 1.9 M \LaTeX\ paper sources and suitable for a wide variety of analyses and NLP applications. We base our approach to data set creation and format on existing works, while also addressing their shortcomings. Improving upon these tried and tested predecessors, unarXive 2022 offers the most complete citation network and most structured content compared to existing data sets, and is surpassed in size only by the PMC-OAS, which covers a different set of disciplines.

With our data set we provide data for two NLP tasks, content based citation recommendation and IMRaD classification, to facilitate its usage. We furthermore distribute our work under consideration of the FAIR principles, sharing it through well established channels and permissively licensed, thereby ensuring proper accessibility, easy use, and possibilities for adaption and extension.

We plan to incrementally update our data set with new arXiv.org submissions. For future developments, we note the importance of mathematical notation in academic publications, as reflected by recent SemEval tasks in 2021 and 2022~\cite{semeval21_task8,semeval22_task12}. Similar to existing projects,\footnote{See \url{https://github.com/PierreSenellart/theoremkb}.} we plan to investigate novel analyses and applications based on the combination of our data set's citation network and structured mathematical notation.

\section*{Author Contributions}  %
Tarek Saier: Conceptualization, Data curation (lead), Formal analysis, Methodology, Software (lead), Visualization, Writing -- original draft, Writing -- review \& editing. Johan Krause: Data curation (support), Software (support). Michael F{\"a}rber: Writing -- review \& editing.

\begin{acks}
This work was partially supported by the German Federal Ministry of Education and Research (BMBF) via [KOM,BI], a Software Campus project (01IS17042).
The authors acknowledge support by the state of Baden-W{\"u}rttemberg through bwHPC.
We thank Johannes Reber for supporting early stages of the software development.
\end{acks}

\bibliographystyle{ACM-Reference-Format}
\bibliography{paper}
\end{document}